\documentclass[amsmath,superscriptaddress,amsmath,amssymb, aps,longbibliography,floatfix,
twocolumn, a4paper]{revtex4-2}
\pdfoutput=1
\usepackage{graphicx}
\usepackage{braket}
\usepackage{dcolumn}
\usepackage{bm}
\usepackage{hyperref}
\usepackage[T1]{fontenc}
\usepackage{xcolor}
\usepackage{subfigure}
\usepackage{appendix}
\usepackage[english]{babel}
\usepackage{natbib}
\usepackage[normalem]{ulem}

\begin{document}
\title{
Dynamical witnesses and universal behavior across chaos and non‑ergodicity in the tilted Bose–Hubbard model
}
\author{Carlos Diaz-Mejia}
\affiliation{Instituto de Ciencias Nucleares, Universidad Nacional Aut\'onoma de M\'exico, Apdo.
Postal 70-543, C.P. 04510  CDMX, Mexico}
\author{Sergio Lerma-Hern\'andez}
\affiliation{Facultad de F\'isica, Universidad Veracruzana, Campus Arco Sur, 
Paseo  112, C.P.
91097 Xalapa, Veracruz, Mexico}
\author{Jorge G. Hirsch} 
\affiliation{Instituto de Ciencias Nucleares, Universidad Nacional Aut\'onoma de M\'exico, Apdo.
Postal 70-543, C.P. 04510  CDMX, Mexico}

\date{\today}
\begin{abstract}
Quantum chaos in isolated quantum systems is intimately linked to thermalization and the rapid relaxation of observables. Although  the spectral properties of the chaotic phase in the tilted Bose-Hubbard model have been well characterized,  the corresponding  dynamical signatures   across the transition to regularity remain less explored
. In this work, we investigate this transition by analyzing the time evolution of the survival probability, the  single-site entanglement entropy, and the half-chain imbalance. Our results reveal a clear hierarchy in the sensitivity of these observables:  the relaxation value of the entanglement entropy varies smoothly as a function of the Hamiltonian parameters  across  the chaos-regular transition,   
while the imbalance exhibits  a more pronounced distinction. Most notably, the survival probability emerges as the most robust indicator of the transition between  chaos and regularity. 
When appropriately   scaled, all  three observables converge onto a common  behavior as a function of the Hamiltonian parameters for different  numbers of sites and bosons,enabling a universal characterization  of the transition between chaotic and regular dynamics. 
\end{abstract}

\maketitle

\section{Introduction}

The question of how closed quantum systems 
thermalize is fundamentally tied to the concepts of quantum chaos and  ergodicity\cite{srednicki1994,Rigol2008,Borgonovi2016,Dalessio2016}.
According to the Eigenstate Thermalization Hypothesis 
(ETH)\cite{Rigol2008}, the eigenstates of chaotic, non-integrable Hamiltonians enable subsystems to thermalize, while integrable systems with extensive conserved quantities evade thermalization.
The tilted Bose-Hubbard model (TBH) has emerged as a paradigmatic system for studying this interplay, as its parameters allow one to tune between integrable and chaotic regimes \cite{Rodr_tilt,buch_03,schneider2019fock}.
Its integrable limits are well-known: localized Fock states for zero hopping ($J=0$), Wannier-Stark states for zero interaction ($U=0$) and momentum-k states for  hopping dominating interaction and tilt terms ($J\gg U,D$)\cite{beringer2024controlling}.
The non-integrable region hosts a rich and tunable chaotic phase.
Recent work has meticulously mapped the static properties of this chaotic phase\cite{Rodr_tilt} 
providing
a comprehensive  characterization of the chaotic regime through the spectral statistics and the multifractal properties of eigenvectors.
However, other fundamental implications of quantum chaos are dynamical.
While spectral analysis identifies the existence of a chaotic regime, it does not directly reveal how the  transition into and out of this regime is experienced by a quantum state evolving in time.
A crucial question is: how does the breakdown  of ergodicity, as one moves from the chaotic phase towards the integrable limits, manifests in the real-time dynamics of experimentally 
accessible observables?
In this work, we fill this gap by investigating the dynamical signatures of the transition from quantum chaos to regularity in the TBH.

We numerically simulate the non-equilibrium quantum quench dynamics from averages over Fock initial states, 
monitoring key observables that are directly probed in cold-atom  experiments\cite{guo2021stark,wang2021stark,moessner_2020,Greiner2002,Lukin_2019}: the survival probability\cite{Haake2010,Torres-Herrera(2017-2),Torres-Herrera(2019),delaCruz_20}, the entanglement entropy, and  the imbalance.


Our results show that these quantities, each with its own sensitivity, act as dynamical indicators of the underlying regular-chaos transition.
We show that when  interaction and tilt are similar  ($U \sim D\sim J$), 
the dynamics exhibit rapid decay of survival probability and fast relaxation of entanglement entropy and  imbalance, 
---all consistent with thermalization in the chaotic ergodic phase.
In contrast, when the tilt 
dominates ($D >>  U$), Wannier-Stark Localization (WSL)\cite{stark_mbl,morong2021observation,therm_vs_bloch_osc_ribeiro_20} takes place, the dynamics are  starkly different, showing  deep oscillations in the survival probability and entanglement entropy and  slow loss of memory of the initial state
---hallmarks of a non-ergodic, regular regime.
By tracking the temporal behavior of these  observables, we dynamically trace the boundary between chaos and regularity, providing a complementary and experimentally actionable  perspective on the phase diagram established by static measures.

This paper is organized  as follows.  Sec.~\ref{sec:model}  introduces the tilted Bose-Hubbard Hamiltonian.  Section~\ref{sec:static}   defines the   observables used in our analysis and 
examines  the 
static properties  of the  Hamiltonian eigenstates across the transition between chaos and regularity.
In Sec.~\ref{sec:dynamics}, we present the distinct dynamical responses  of these observables as the system moves  from the chaotic regime  toward the interaction-dominated  and  tilt-dominated limits, and we  characterize the dynamical signatures of  this transition using  averages over  Fock initial states.
Finally, we present our conclusions in 
Sec.~\ref{sec:conclusions}.

\section{The model}\label{sec:model}
We analyze the one-dimensional tilted Bose-Hubbard model with $N$ bosons on $M=N$ sites and open boundary conditions\cite{Jaksch1998,Rodr_tilt,
schneider2019fock,kolovsky2003new},
\begin{equation}
    H = -J \sum_{i=1}^{M-1} (\mathbf{b}_i^\dagger \mathbf{b}_{i+1} + \text{h.c.}) + \sum_{i=1}^M \left[ \frac{U}{2} \mathbf{n}_i(\mathbf{n}_i-\mathbf{1}) + D\, i\, \mathbf{n}_i \right]
\end{equation}
where $\mathbf{b}_i^\dagger$ ($\mathbf{b}_i$) are the bosonic creation (annihilation) operators at site $i$, and $\mathbf{n}_i = \mathbf{b}_i^\dagger \mathbf{b}_i$ is the number operator. The  Hilbert space dimension  is $\mathcal{D}(N,M)=(M+N-1)!/(N!(M-1)!)$.
The model  is governed  by three parameters:  the  tunneling amplitude $J$, the on-site interaction $U$,  and  the tilt $D$.
Because  the  tilt breaks  the parity symmetry, particle number is   the only remaining conserved quantity.

Throughout  this work   we set  $J=1$ as the unit of energy  and explore the model by varying  $U$ and $D$. We characterize  the transition from integrability to quantum chaos using  both local and non-local  dynamical observables. A recent detailed study has analyzed the interplay of these parameters  using level statistics  and multifractal analysis\cite{Rodr_tilt}.
As noted  above, the system possesses  integrable limits: when $U \rightarrow 0$, the system admits  analytical solutions known as Wannier-Stark states and  dynamically  exhibits Bloch oscillations \cite{meinert2014interaction,hartmann2004dynamics}.
 When $D\rightarrow0$, one recovers the standard Bose-Hubbard system\cite{pausch2022optimal,delaCruz_20}, which itself has two integrable limits.  For $U\gg J$, the system maps onto   free fermion  (the Hard-Core Boson limit, HCB). For $J\gg U$, the system  becomes that of   free, non-interacting bosons in a  tight-binding chain, diagonal in momentum space.
 
 We investigate  the behavior of the model   as the  tilt  is increased (at fixed interaction), tracing the progression  from the chaotic Bose-Hubbard  regime through the  chaotic tilted Bose-Hubbard regime and finally into the  integrable Wannier-Stark Localization limit---that is,  the cBH-cTBH-WSL transition (vertical line in Fig.\ref{fig:rcaos}). Complementarily,  we  examine  the path where  the  interaction strength is increased (at fixed tilt), driving the system from the integrable Wannier-Stark-localization regime, through  the  chaotic Tilted Bose-Hubbard regime and   ultimately  into the   integrable Hard-Core Bosons limit, corresponding to   the WSL-cTBH-HCB transition (horizontal line in Fig.\ref{fig:rcaos}).

\section{Static signatures of the transition chaos to regularity. 
}
\label{sec:static}

\subsection{Quantum chaos}
We adopt the standard  quantum-chaos  criterion based on the similarity between  eigenenergy correlations and those of an appropriate  random-matrix ensemble, which for the present model corresponds to  the Gaussian Orthogonal Ensemble (GOE).  
Quantum chaotic systems exhibit unfolded level spacings $\tilde s_n=(E_{n+1}-E_n)\rho(\bar{E}_n)$, where $\rho(E)$ is the density of states and $\bar{E}_n=(E_{n+1}+E_n$)/2, that follows the Wigner-Dyson distribution, whose main characteristic is  level repulsion ($\lim_{\tilde s \rightarrow 0} P(\tilde s)=0$) \cite{Haake2010,Borgonovi2016}.
To avoid the unfolding procedure, one can instead  use the mean ratio of  consecutive level spacings \cite{Oganesyan_2007}\cite{Huse_2015}
\begin{equation}\label{ratio1}
\langle r\rangle=\left \langle \frac{\min(s_n, s_{n-1})}{\max(s_n, s_{n-1})} \right \rangle,
\end{equation}
where $s_n = E_{n+1} - E_n$. This ratio is independent of the  density of states, making   unfolding unnecessary.
For quantum-chaotic systems, the mean level spacing ratio is $\langle r \rangle_W  \approx 0.535$, while  regular spectra yield the  lower  value  $\langle r \rangle_P \approx 0.386$.

\begin{figure}
    \centering
    \includegraphics[width=0.99\linewidth]{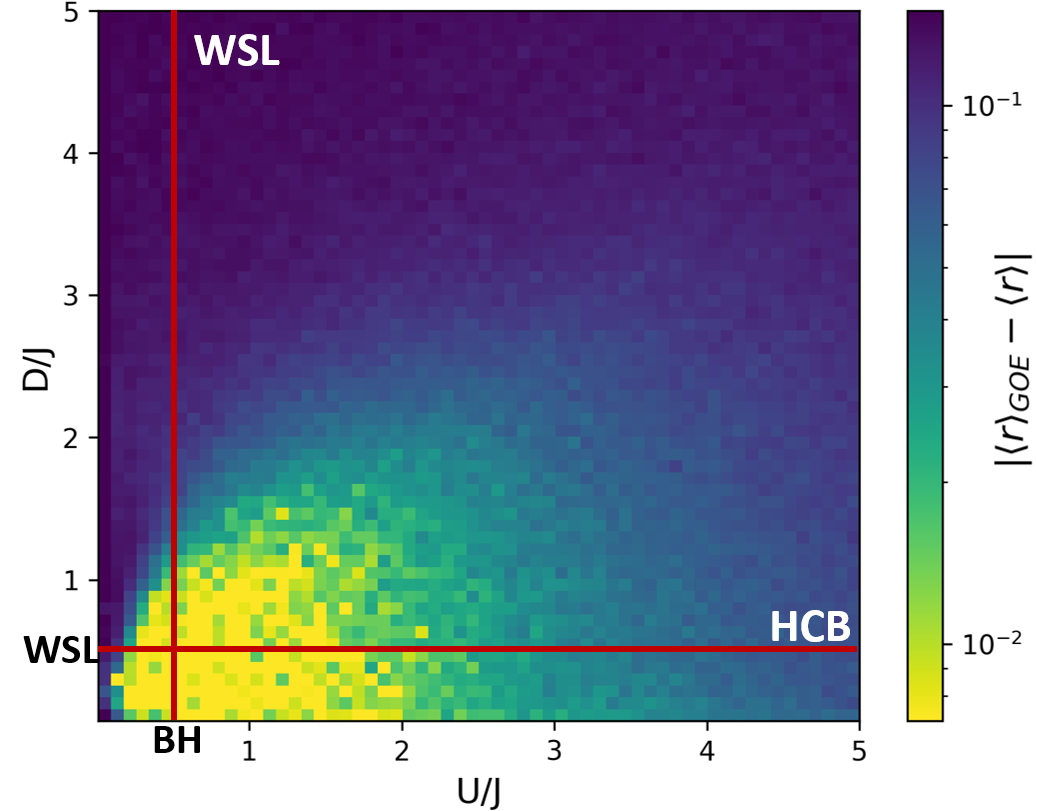}
    \caption{Absolute value of the difference between the mean level-spacing ratio $\braket{r}$ of the eigenspectrum of  the tilted Bose-Hubbard  model  and the chaotic GOE prediction, shown  as a function of the interaction  $U/J$  and the  tilt 
    for a system with $N=M=8$. The red lines indicate the parameter paths  explored in the numerical simulations. The acronyms indicate   the following limits: WSL (the integrable Wannier-Stark Localization ), HCB (the integrable Hardcore Bosons), and BH (the chaotic Bose-Hubbard model at $D=0$). 
    }
    \label{fig:rcaos}
\end{figure}

Our analysis of the mean level spacing ratio, $\langle r \rangle$,  reveals a rich structure in the chaos-regularity phase diagram  as a function of interaction strength $U$ and tilt $D$ (Fig. \ref{fig:rcaos}). For very small tilt $D\approx 0$,  the symmetry sectors of the non-tilted Bose-Hubbard model are broken, and the mean spacing ratio $\braket{r}$ exhibits  GOE behavior for $U/J\in[0.1,2]$.  This  agrees  with known  results for the  non-tilted  Bose-Hubbard model,  where each parity sector displays GOE correlations over the same interval\cite{pausch2022optimal}. 
A similar breaking  of  parity symmetry  occurs in the interacting Aubry-André model when a very weak  onsite disorder is introduced\cite{diaz2024persistent}.

 The onset of chaotic behavior shows  a pronounced asymmetry in  the systems' response to the  two parameters:   chaos is rapidly suppressed as  $D$ increases at fixed $U$, whereas it remains much  more robust when  $U$ is increased at fixed $D$.
As we demonstrate below,  this asymmetry  is  also reflected  in the  dynamical observables.

\subsection{Participation ratio}

The participation ratio (PR) of a normalized state $\ket{\Psi} = \sum_k c_k \ket{k}$ expanded in an orthonormal   basis $\{\ket{k}\}$, is defined as 
\begin{equation}
    \text{PR}=\frac{1}{\sum_k |c_k|^4}.
\end{equation}

The PR quantifies   how delocalized  the state $\ket{\Psi}$ is within the 
chosen basis.
When a state $\ket{\Psi}$ is evenly  distributed in the basis,  the PR is as large as the dimension ($\mathcal{D}=\mathcal{D}(N,M)$) of the Hilbert space, whereas  a state  localized in a single basis  element  has   $\text{PR}=1$. In chaotic regimes where eigenstates resemble those of the   Gaussian orthogonal ensemble,  one expects $\text{PR}\approx\mathcal{D}_{\text{GOE}}= \mathcal{D}/3$ \cite{Kaplan_2007}.

\begin{figure}
    \centering
    \includegraphics[width=0.49\linewidth]{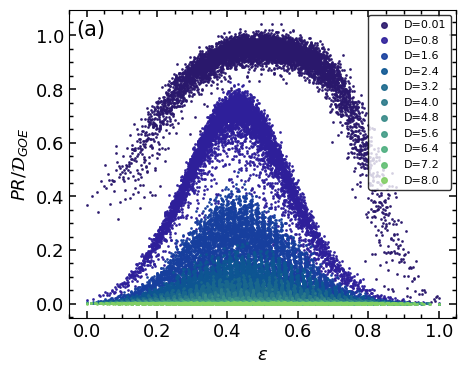}
    \includegraphics[width=0.49\linewidth]{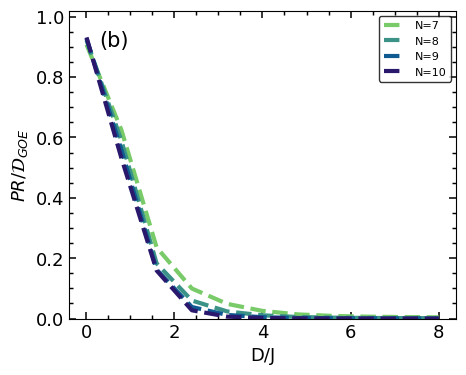}
    \includegraphics[width=0.49\linewidth]{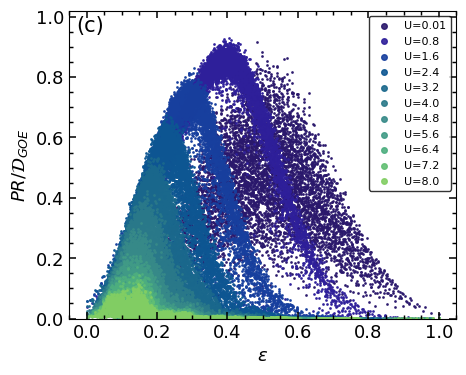}
    \includegraphics[width=0.49\linewidth]{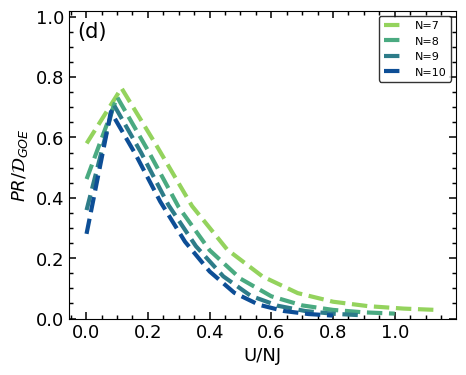}
    \caption{Left: $PR/\mathcal{D}_{GOE}$ of the eigenstates in the Fock basis for $N=8$. Right: scaling analysis averaging over $ 80\%$ of the eigenstates in the center of the spectrum.
In the  upper panel, interaction is fixed at $U=0.5J$ and $D$ varied; in the lower panel, tilt is fixed at $D=0.5J$ and $U$ varied.
}
    \label{fig:Prs_eigenvecs}
\end{figure}

To allow a consistent comparison across systems with different tilts and interaction strengths, we present our results in terms of the normalized energy,
\begin{equation}
    \epsilon=\frac{E-E_{min}}{E_{max}-E_{min}}.
\end{equation}

Fig. \ref{fig:Prs_eigenvecs} shows how the  participation ratios of the eigenstates in the Fock basis change  as  the interaction or tilt varies.
In  panel (a),  for  very small  tilt ($D=0.01$), chaos arises throughout  the spectrum,  and the participation ratio behaves as a smooth function of energy with low dispersion of values. Only the eigenstates at the edges of the spectrum ---those with very high and  very low energy--- display  low small PR values, indicating  stronger  localization in the Fock basis. As the tilt  increases, the PR decreases uniformly across the spectrum, consistent with a system 
entering the  Wannier-Stark-Localization regime. This trend is also reflected in the  scaling analysis shown in panel (b).  It is worth noting  that the larger the system,  the more  sensitive it becomes to the   Wannier-Stark-Localization (WSL) transition as  the tilt  increases.

\begin{figure}
    \centering
    \includegraphics[width=0.49\linewidth]{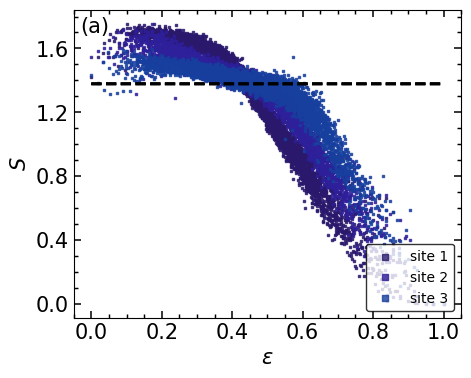}
    \includegraphics[width=0.49\linewidth]{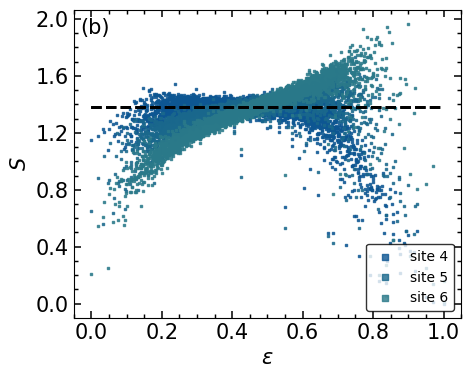}
    \includegraphics[width=0.49\linewidth]{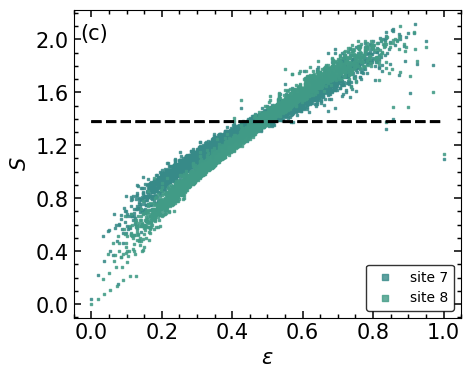}
    \includegraphics[width=0.49\linewidth]{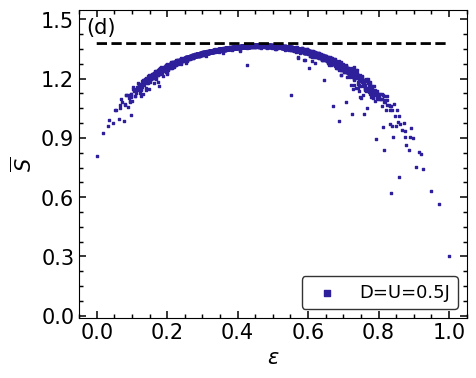}

    \caption{Single-site entanglement entropy of eigenstates for $N=8$.
Panels (a)--(c) show $S^{(i)}$ grouped by sites (from edge to middle), revealing how spatial position and energy density shape the entanglement landscape;
panel (d) displays the site average $\overline{S}$. The black dashed line is the Page value. Parameters are chosen in the chaotic window, where most mid-spectrum eigenstates are near-thermal yet site-dependent deviations persist.}
    \label{fig:entropies}
\end{figure}

Panel (c) of  Figure~\ref{fig:Prs_eigenvecs}, reveals   a different behavior    when the interaction strength is varied.
Two integrable limits are expected: 1) when the interaction approaches zero (Wannier-Stark eigenstates), the PR exhibits  a wide dispersion of values across all the energy range, and 2) when interaction dominates, we have the  HCB integrable regime. Within  the chaotic regime ($U\in[0.8,2]J$),  the PR  behaves  smoothly as a  function of energy.
The PR of eigenstates  with higher energy rapidly decay, while the eigenstates in the middle of the spectrum  present values of PR  organized in a narrow band  extending through the low-energy region of the spectrum.
The scaling analysis in panel (d) shows   that adding more bosons does not qualitatively change this trend, but the transition toward the HCB integrable  regime becomes 
more pronounced as  the number of bosons increases.

A noteworthy observation from the scaling analysis in the   right panels of Fig~\ref{fig:Prs_eigenvecs} is the striking similarity  of the curves across all examined system sizes. This collapse  is achieved  by appropriately scaling  the variables: the participation ratio is normalized  by the GOE reference value  $\mathcal{D}_{GOE}=\mathcal{D}/3$, and the interaction strength is rescaled  by the number of bosons $N$,  consistent with the two-body nature  of the interaction term.  It is also interesting that the participation ratio of the eigenstates does not reach the  GOE expectation, i.e.,  $\text{PR}/\mathcal{D}_\text{GOE}<1$.

\begin{figure}
    \centering
    \includegraphics[width=0.49\linewidth]{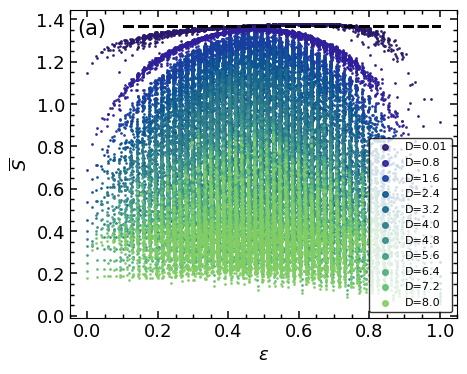}
    \includegraphics[width=0.49\linewidth]{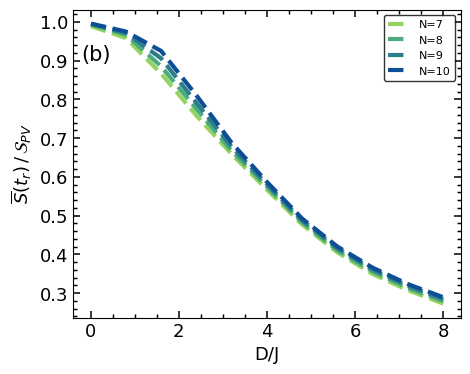}
    \includegraphics[width=0.49\linewidth]{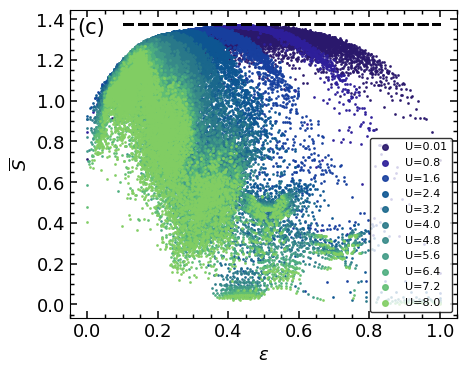}
    \includegraphics[width=0.49\linewidth]{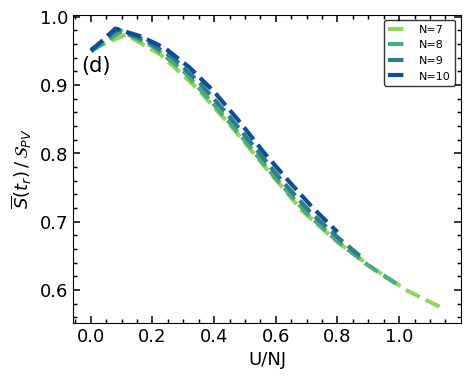}
    \caption{Left: $\overline{S}$ of the eigenstates in the Fock basis for $N=8$, the black dashed line is the Page value $\mathcal{S_P}$. Right:  $\overline{S}/\mathcal{S_P}$ scaling analysis averaging over $ 80\%$ of the eigenstates in the center of the spectrum .
In the  upper panel, interaction is fixed at $U=0.5J$; in the lower panel, tilt is fixed at $D=0.5J$
Single-site entanglement entropy of eigenstates for $N=8$. }
    \label{fig:S_eigenvecs}
\end{figure}

\subsection{Entanglement Entropy}

The entanglement entropy of eigenstates is a widely used observable~\cite{stark_mbl,wang2021stark} for characterizing thermalization \cite{Greiner2002}, quantum chaos\cite{piga2019quantum}, non-thermal behavior such as many-body scars\cite{scars_bose_sim}, and many-body localization transitions\cite{Lukin_2019}.
In this section, we employ it to analyze the crossover between integrability and chaos, first for the eigenstates and later  
in Sec.\ref{sec:dynamics} for the dynamics of initial Fock states.

We consider the entanglement entropy between a single site and the rest of the chain.  Since  we are considering a tilted non-periodic  chain, each site is inequivalent  and no site is  preferred. 
To obtain  a consensus indicator of chaotic versus regular behavior, we  average the entanglement entropy  over all single-site subsystems. It is important to note, however,  that the entanglement entropy of different  individual sites conveys complementary information.

For any site $i$, the single-site entanglement entropy is  defined as

\begin{equation}
    S^{(i)} \;=\;
-\,\mathrm{Tr}\!\left[\rho_i \ln \rho_i\right],\qquad \rho_i=\mathrm{Tr}_{\bar{i}}\,\rho,
\end{equation}
where the second trace is taken over the complement of site $i$ (the remaining $M\!-\!1$ sites, denoted as $\bar{i}$) and $\rho$ is the pure-state density matrix of the entire system.
The average over all sites is

\begin{equation}
\overline{S}=\frac{1}{M}\sum_{i=1}^M S^{(i)}\,.
\end{equation}

When this average is taken, the expected typical entanglement entropy for chaotic states   in  systems with conserved particles was calculated in Ref.\cite{yauk2024typical} as   \begin{equation}\mathcal{S}_{PV }=Vf[(n+1)\ln (n+1)-n\ln n]+1/2(f+\ln (1-f)),
\label{eq:TheoreticalPV}\end{equation}
where $f=1/M$ is  the subsystem fraction, $n=N/V$ is the  particle density  and $V=M$ is the volume of the system. We  refer to this quantity  as the Page value. $\mathcal{S}_{PV}$ serves   as an upper bound for the averaged entanglement entropy and provides  
 an important  reference value  for  the  following analysis.

In Fig.~\ref{fig:entropies} we examine  how the entanglement entropy of single sites varies    from the bottom site $S^{(1)}$ to the highest site $S^{(M)}$, for $M=N=8$ in a chaotic regime with $U/J=D/J=0.5$.
 Sites are grouped into  three sets: in  panel(a) the  first three sites display    very similar entanglement,  reflecting boundary effects. High-energy eigenstates are less likely to exhibit large entanglement. In panel (b) the entanglement entropy of sites 4,5 and 6  exhibits a reduced dispersion of  values for states near  the center of the spectrum. Nevertheless, some of these eigenstates appear
as athermal outliers of 
 ETH expectations. In panel (c) the entanglement entropy of  sites 7 and 8 again reveals   boundary  effects,  with  high-energy states exhibiting  large entanglement.
Finally, in panel (d) the average over all sites shows very low dispersion, consistent with ETH.    At low and high energies, the averaged entropy is reduced  compared to the central region, where it approaches  the Page value $S_{PV}$. 
Yet,  athermal outliers remain visible in the chaotic center of the spectrum. 

The averaged one-site entanglement entropy  provides   a useful  diagnostic of  the transition between regular and chaotic regimes. 
The  upper panels of Figure  \ref{fig:S_eigenvecs} illustrate how  this transition evolves  as the  tilt strength increases from weak to strong. In panel  (a), we present the averaged entanglement entropy of all eigenstates for a fixed interaction strength $U=0.5 J$ and varying tilt $D$. At $D=0.01$, the system  is close to  the chaotic Bose Hubbard limit; correspondingly, the  entanglement entropy is high for most eigenstates and exhibits  low dispersion. As the  tilt  increases ($D=0.8$ and $D=1.6$), eigenstates away from the center of the  spectrum  begin to show  entropy values  below the Page value $\mathcal{S}_{PV}$. For even  larger tilt values, the  entanglement entropy  decreases  
uniformly,  and degeneracies in the energy spectrum  become evident. Nevertheless, unlike the participation ratio,   entanglement entropy does not  exhibit an abrupt change at higher tilt strengths.  

The previous  behavior is highlighted by comparing  panel (b) of Fig.\ref{fig:S_eigenvecs} with the corresponding  panel (b) of Fig.~\ref{fig:Prs_eigenvecs}. In Fig.\ref{fig:S_eigenvecs}(b),  the mean entanglement entropy of  eigenstates in the middle of the spectrum is plotted   as a function of the ratio $D/J$ for  system sizes $N=M=7,8,9$ and $10$.  Remarkably,  when  the mean entanglement entropy is rescaled by  the corresponding Page value $\mathcal{S}_{PV}$,  the resulting curves collapse  onto one another, forming  a  nearly universal curve that is largely  independent of system size.

On the other hand, panel (c) of Fig \ref{fig:S_eigenvecs} illustrates the effect of varying the interaction strength at a fixed tilt. In this case,  higher- energy  eigenstates are markedly  more  sensitive to   changes in the  interaction, as evidenced by a  rapid decrease in their entanglement  entropy. For weak  interactions ($U=0.01, 0.8$ and $1.6$),  the entropy values form   an  arch-like structure; broad and highly dispersed  at $U=0.01$,  and   narrowing  at $U=0.8$. This is   consistent with the system residing in the regular   WSL  regime at very low interaction strengths. As the interaction strength increases further    and the system crosses into   the regular  HCB regime ($U\geq 2.4$), the arch-shaped  structure disappears  and  entropy values become significantly    more dispersed. However, unlike    the  PR  shown in panel (b) of Fig.\ref{fig:Prs_eigenvecs},  the  entanglement entropy does not drop to   values substantially below  the corresponding reference value (here $\mathcal{S}_{PV}$, and   $\mathcal{D}_{GOE}$  in the case of the PR). 

In panel (d) of Fig.\ref{fig:S_eigenvecs},  we present  the   mean entanglement entropy of eigenstates in the center of the spectrum as a function of the rescaled interaction parameter [$U/(NJ)$] for different system sizes. The mean entropy is normalized  by the corresponding $\mathcal{S}_{PV}$, yielding  curves that  collapse across all  all four system sizes considered. As expected, the maximum of this nearly  universal curve does not occur at  $U=0$, but  rather around   $U/(NJ)\approx 0.15$, in  agreement with the chaotic region identified in  Fig.\ref{fig:rcaos}. 

\subsection{Imbalance }
Another well-known observable is the imbalance~\cite{stark_mbl,mbl_yao_2020,bloch_osc_mbl_refael}. As a complement to the entanglement entropy, the imbalance tracks how the boson population distributes across the chain. It is also far more accessible experimentally and  effectively distinguishes between  chaotic and localized phases.
In spin and fermionic  systems, the  odd--even imbalance is commonly used \cite{scherg2021observing,wu2019bath,bloch_osc_mbl_refael}.
Here, instead,  we employ  the half-chain imbalance, defined as
\begin{equation}
    \mathcal{I}=\frac{\braket{n_{l}}-\braket{n_{r}}}{N},
\end{equation}
where $n_l = \sum_{i=1}^{\lceil M/2\rceil } n_i$ and $n_r = \sum_{i=\lceil M/2\rceil +1}^{M} n_i$ are the total particle numbers in the left and right halves of the chain, respectively, and $\lceil x\rceil$ denotes  the nearest integer that is larger than or equal to $x$. When the number of sites $M$ is  even, both  halves have the same  length, while for odd $M$  the right half contains  one site fewer than the left.

\begin{figure}
    \centering
    \includegraphics[width=0.49\linewidth]{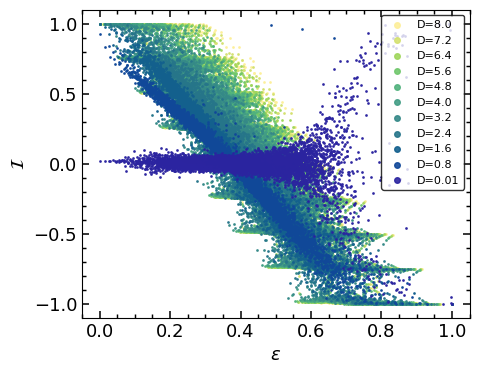}
    \includegraphics[width=0.49\linewidth]{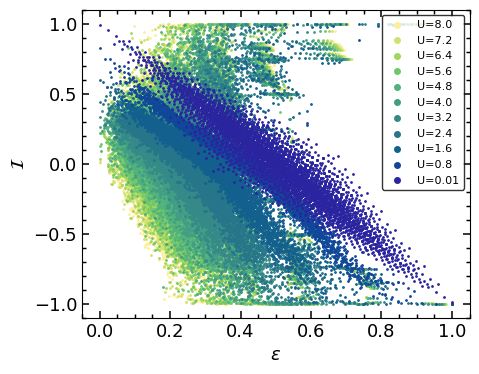}

    \caption{Half-chain imbalance for eigenstates for different tilt  strengths at fixed interaction $U=0.5J$ (left) and  for varying  interaction strength at fixed tilt $D=0.5J$  (right).
As $D/J$ increases, eigenstates cluster near specific imbalance values, forming a ladder-like structure characteristic of Wannier-Stark Localization tendencies.
}
    \label{fig:Imb_Ds_Us}
\end{figure}

Figure \ref{fig:Imb_Ds_Us} shows   the imbalance   as a function of the rescaled energy of the Hamiltonian eigenstates.  In panel (a), for $D=0.01J$,  the imbalance is narrowly concentrated  around zero with small  dispersion, except for  the highest-energy states. As the tilt  increases slightly, the imbalance values spread diagonally  across the  full   interval $[-1,1]$: high-energy states occupying the right half of the chain, while  low energy eigenstates occupying the left half.  For small  tilt, the imbalance  remains     a smooth function of energy. With further increase in  tilt,  the diagonal pattern persists but becomes more dispersed, signaling  a transition  to a non-chaotic regime.  At very large tilt, 
the imbalance organizes  into  discrete  values, forming  a ladder-like structure  characteristic of Wannier-Stark localization. 


 Panel (b) displays  the imbalance  for different interaction strengths at  fixed tilt  $D=0.5 J$. Because the  tilt is already large, the imbalance exhibits   a diagonal pattern  for small   interactions. In this parameter region --- where the spectral statistics indicate chaotic behavior ---the imbalance shows low   dispersion and  varies    smoothly with energy.  As the  interaction increases, the spectrum compresses toward low energies, 
 and the imbalance values  spread vertically, breaking the diagonal structure. 
Similar  to the tilt-dominated regime, some  states  cluster around specific imbalance values; however, most  eigenstates display  values with no clear  pattern.
\section{Dynamical signatures of transition  chaos to regularity }\label{sec:dynamics}

Dynamical properties are equally revealing of the regular-to-chaotic crossover. We illustrate  this by examining  the behavior of  three observables as functions of either the interaction strength  $U$ or the  tilt $D$, while  keeping the other  fixed and varying it  until one of them dominates.
The  observables we consider  are 1) the survival probability, a global measure that directly reveals long-range spectral correlations through the so-called correlation hole; 2) 
the on-site entanglement entropy, which grows ballistically and relaxes rapidly in chaotic regimes~\cite{kim2013ballistic,piga2019quantum};
and 3) the half-chain imbalance, which also increases or decays to zero very quickly when chaos is prevalent.
As  the system approaches regularity, these behaviors change significantly: the survival probability lacks the  distinctive  correlation hole and shows strong initial  fluctuations;
 entanglement growth becomes logarithmic~\cite{serbyn2013universal}; and the imbalance remains close to its initial value, consistent with WSL behavior.


We analyze  the dynamics of these observable using  sets of initial Fock product states   with on-site occupations restricted to   $n_i\leq 3$ for all $i=1,...,M$. Such states are experimentally realistic  and have been realized in recent setups (e.g., half filling or two bosons per site) \cite{Lukin_2019,scherg2021observing,scars_bose_sim}. 
For the survival probability, and for each of the  four  system sizes considered,    we randomly  select  200 Fock states from those whose  mean energies lie within  the energy interval  $[E_c-0.4 \Delta E_s, E_c+0.4 \Delta E_s]$, where $E_c$ and $\Delta E_s$ are the mean energy and standard deviation, respectively, of the energy spectrum   corresponding to  the parameters $U=0.5J$ and $D=0.8 J$ ---a point  well inside   the chaotic regime according to the map in Fig.\ref{fig:rcaos}. This same set of initial  Fock states is used throughout our  analysis of the survival probability 
 when  the Hamiltonian parameters are varied. 

For the entanglement entropy, we follow an analogous selection  procedure.  The only  difference is  that, due to numerical limitations,  we use fewer initial  states: specifically,    $50$ random   Fock states for system sizes $N=M=7,8$ and $9$,  and $20$  states  for the largest system  $N=M=10$.

For the imbalance,  we choose   initial Fock states  that not only satisfy  $n_i\leq 3$ but also  have an initial  imbalance $\mathcal{I}=-1$.  These  correspond to    states with all bosons on the right half of the chain and none on the left.  For system sizes $N=M=7,8$ and $9$,  we use all  Fock states satisfying  these conditions, which amount to  6, 31 and 20 states, respectively. For the largest system size, $N=M=10$, the number of qualifying  states increases to  101; however,  due to numerical constraints,   we randomly select only 20 states of them.  

In the  Appendix,  we present  for the case $N=M=8$  the complete set of $50$ states used for  the entanglement entropy analysis  and the set  used for  the imbalance dynamics.

\subsection{Survival Probability}
\begin{figure}
    \centering
    \includegraphics[width=0.9\linewidth]{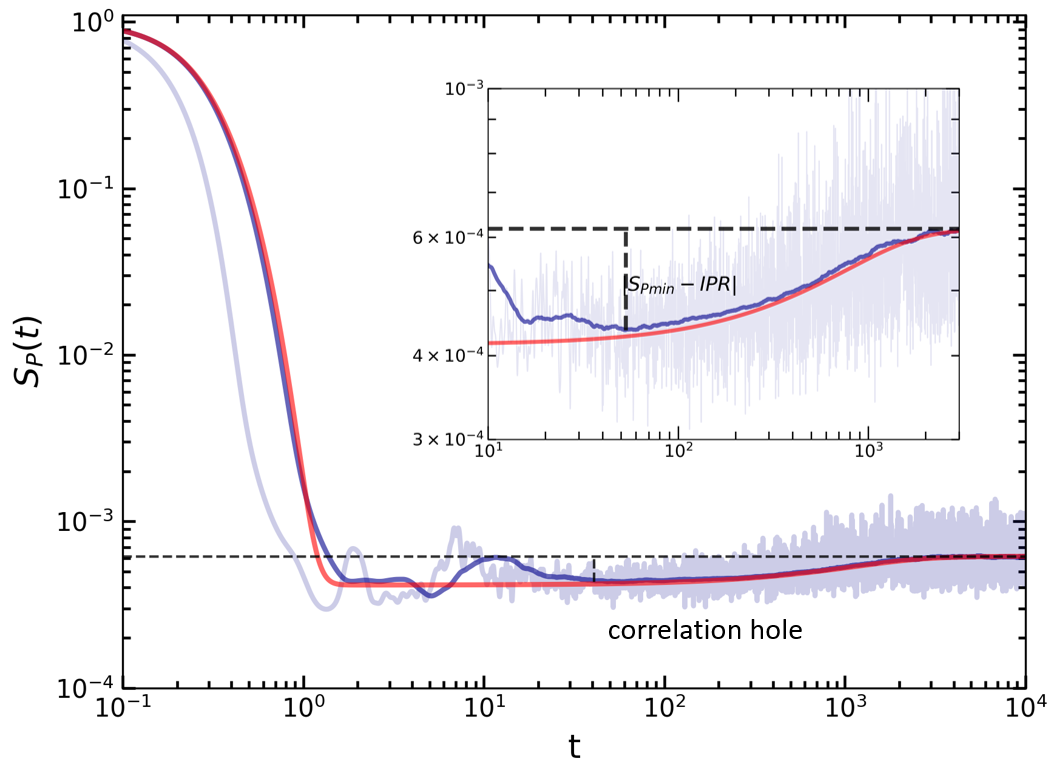}
    \caption{Averaged Survival Probability for an ensemble of 200 Fock states (light blue line) near the center of the spectrum. The inset shows the beginning and end of the correlation hole, for $N=8$, the Hamiltonian parameters are in the most chaotic region $U=D=0.5J$. The dark blue line is a moving average, the red line is the analytical curve averaged over those 200 Fock states, the horizontal dashed line is the relaxation value or IPR and the vertical dashed line in the inset is the depth of the correlation hole respect to the IPR}
    \label{fig:sp_caos}
\end{figure}

The survival probability $S_P(t)$ (or fidelity) is the probability of finding the system in  its initial state $\ket{\Psi(0)}$ after unitary evolution to $\ket{\Psi(t)}$,
\begin{equation}
    S_P(t) \;=\;
\left|\braket{\Psi(0)|\Psi(t)}\right|^2 \;=\;  \left|\sum_m |c_m|^2 e^{-iE_m t} \right|^2,
\label{eq:SP1}    
\end{equation}
where $c_m$ are the components of the initial state in the Hamiltonian eigenbasis, $\ket{\Psi(0)}=\sum_m c_m \ket{E_m}$.
The survival probability is closely related to the spectral form factor (SFF)~\cite{Brezin1997}.
It is convenient to decompose $S_P(t)$ into diagonal and off-diagonal contributions,
\begin{equation}
    S_P(t) \;=\;
\sum_m |c_m|^4\;+\;
\sum_{k\neq m} |c_m|^2|c_k|^2\,e^{-i(E_m-E_k) t}\,.
\end{equation}
At long times, and in the absence of degeneracies,  the temporal average of $SP(t)$ approaches  the inverse participation ratio of the state in the energy eigenbasis, $\overline{S_P}=\mathrm{IPR}=\sum_m |c_m|^4$.
In the chaotic region, $S_P(t)$ develops a pronounced dip before reaching the IPR plateau.  This correlation hole  is a  hallmark of  long-range spectral correlations.
As a specific instance  of the SFF,  the survival probability exhibits the standard dip--ramp--plateau structure, which can be described analytically~\cite{Alhassid92},
\begin{equation}
    \braket{S_P(t)} \;=\; \frac{1-\mathrm{IPR}}{\eta -1}\Big[\eta\, S_P^{bc}(t) \;-\; b_2\!\left(\frac{t}{2 \pi \bar\nu}\right)\Big] \;+\;
\mathrm{IPR},
    \label{eq:S_p-analitical}
\end{equation}
where $S_P^{bc}(t)$ is  the squared modulus of the  Fourier transform of the smoothed local density of states,
\,$S_P^{bc}(t)=\big|\int dE\, \rho(E)\, e^{-i E t} \big|^2$, $\bar{\nu}$ is the mean density of states in the energy region where the dominant components $c_m$ lie,  $\eta=(\int dE\, \rho^2(E)/\nu(E))^{-1}$  quantifies the effective number of   levels involved  and $b_2(\tau)$ is the GOE two-level form factor.
The survival probability has 
emerged as a practical quantum indicator of chaos, since for random initial states its late dynamics  can be directly  related   to the GOE  two-level form factor~\cite{Torres-Herrera2015,Torres-Herrera(2017-2),Torres-Herrera(2019),delaCruz_20}.
 Figure \ref{fig:sp_caos} shows an  example of a typical survival-probability curve in a chaotic regime  along with the corresponding  analytical prediction. The correlation hole appears at middle times ($t\in[20,10^3]$).   The expected value of the survival probability at the minimum  of the correlation hole can be estimated using GOE matrices \cite{schiulaz2019thouless}, yielding $S_{Pmin}\approx 2/\mathcal{D}$. Combined    with the  GOE estimate for the  IPR, $\text{IPR}=1/\text{PR}=3/\mathcal{D}$, we obtain  an estimate   of the depth of the correlation  hole   $|S_{Pmin}-IPR|\approx 1/\mathcal{D}$. Since   this value is very  small, it is more convenient   to characterize  the depth of the correlation hole through  the difference between   inverse quantities $|\frac{1}{S_{Pmin}}-\text{PR}|$. For GOE matrices, this difference is $|\frac{1}{S_{Pmin}}-\text{PR}|\approx \mathcal{D}/6=\mathcal{D}_\text{GOE}/2 $.   This alternative  measure  will  prove  useful as a dynamical witness to distinguish  chaotic from  integrable regimes.

As we will show, the survival probability --- and particularly the depth of its correlation hole---   is the clearest  and most robust  dynamical indicator of the  transition between regularity and chaos. 
We compare the behavior of the correlation-hole depth  with that  of  two local dynamical observables, 
the entanglement entropy and the  imbalance, in order to obtain  a reliable dynamical witness of the  regular-chaos transition.

\subsection{Dynamics of the survival probability, entanglement entropy and imbalance}
In this section, we analyze chaotic and regular dynamics by  studying the evolution of the survival probability, entanglement entropy and imbalance as a functions of the interaction strength $U$ or tilt $D$, for four different system sizes. Despite their distinct physical nature, all three  observables display qualitatively   similar behavior across the  transitions between   regular and chaotic regimes.

\begin{figure}
    \centering
    \includegraphics[width=0.48\linewidth]{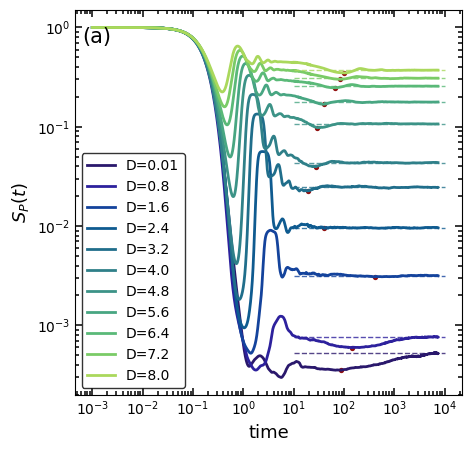}
    \includegraphics[width=0.48\linewidth]{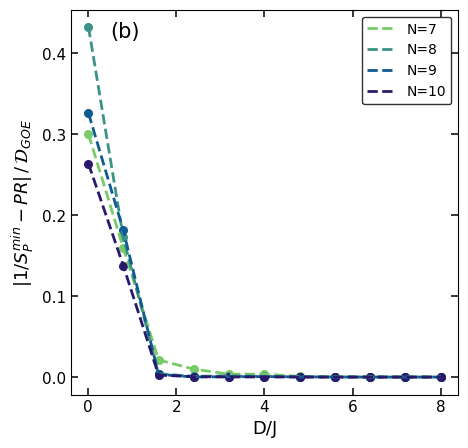}
    \includegraphics[width=0.48\linewidth]{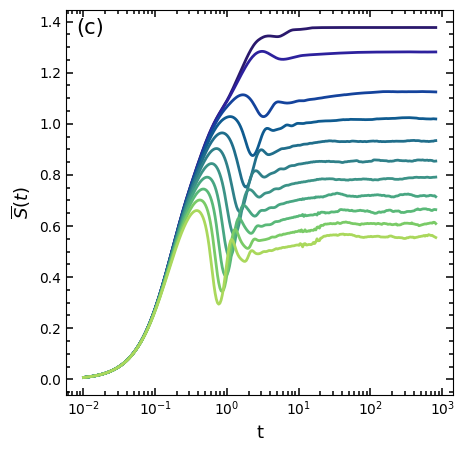}
    \includegraphics[width=0.48\linewidth]{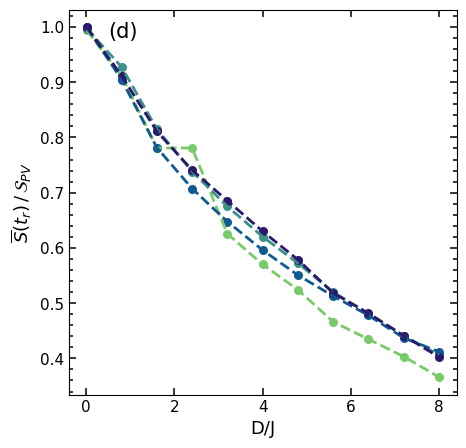}
    \includegraphics[width=0.48\linewidth]{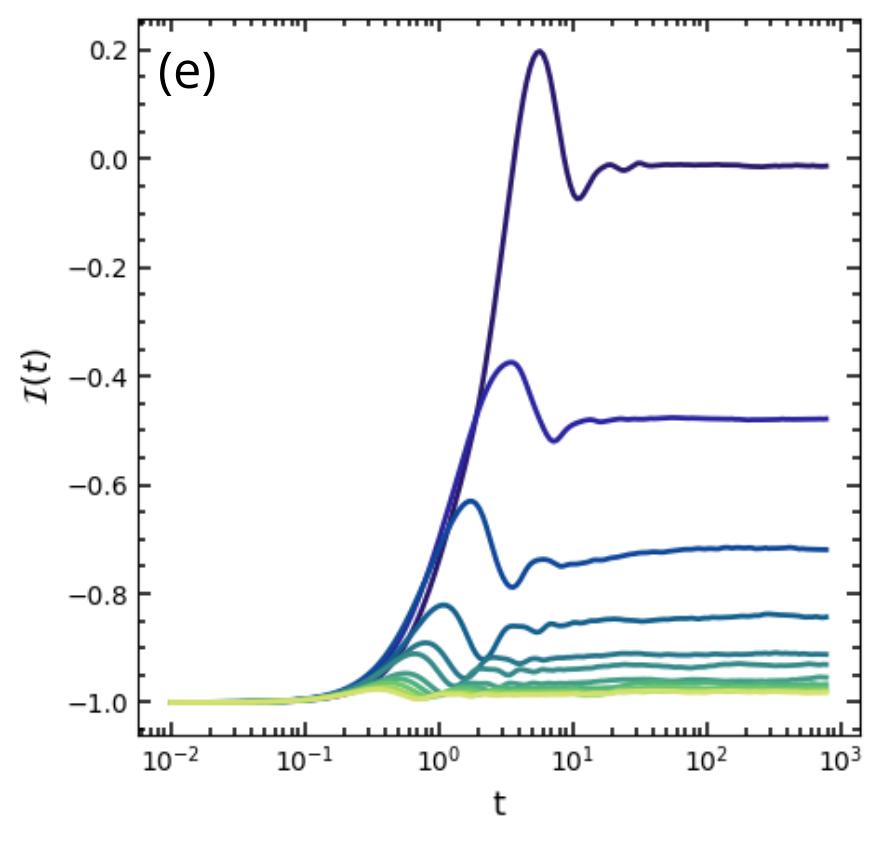}
    \includegraphics[width=0.48\linewidth]{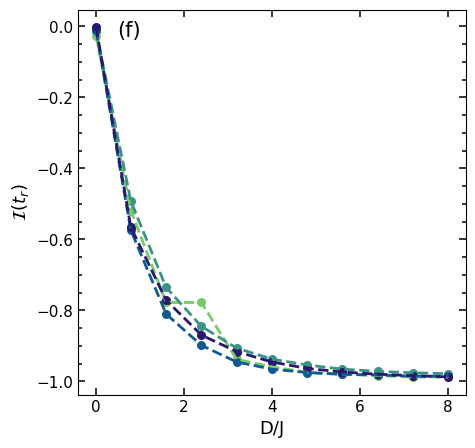}
    \caption{Averaged evolutions of (a) Survival probability $S_P(t)$, (c) entanglement entropy and (e) half chain imbalance for $N=8$ and fixed interaction $U/J=0.5$. Each curve is the average over 50 Fock states (Survival Probability averaged over 200 states) at the center of the spectrum with states initially prepared with $n_{max}=3$ bosons per site for different values of the tilt $D$.
    In the case of imbalance, we chose states with the same initial imbalance $\mathcal{I}(t_0)=-1$, which corresponds to 0 bosons to the left and $M$ to the right part of the chain. All the curves were smoothed with a moving average. On the right panel: (b), (c), (d) we show the relaxation values (averaged over the last 10 values of the evolution) of the observables for different scales; in the case of the survival probability we show the depth of the correlation hole for different scales by scaling with the participation ratio.}
    \label{fig:sps_Ds}
\end{figure}

\subsubsection{Dynamics under tilt variation}
The left panels of Fig.~\ref{fig:sps_Ds}  present the dynamics of the three observables as the  tilt  is varied: the  survival probability, the averaged one-site entanglement entropy and the half-chain imbalance. For each observable, the time intervals shown    were chosen such that  equilibration is reached.  

Panel (a) displays  the  evolution of the survival probability  $S_P(t)$ in log-log scale and  reveals two key  features. First, for  small $D/J$ (the first two curves), a correlation hole develops at intermediate times $t \in [20,10^3]$. This feature disappears as the tilt  increases and becomes fully suppressed for  $D\geq1.6J$. Second, as  the tilt grows, a pronounced  revival emerges after   the initial quadratic decay.
We attribute these revivals   to  Bloch oscillations, which dominate  the dynamics once the tilt becomes sufficiently strong~\cite{meinert2014interaction}. 

The entanglement entropy,  plotted in panel (b), also exhibits  these oscillations  following  its initial rapid  growth.  Because the vertical axis  uses a  linear scale, the strengthening of the oscillations with increasing tilt  becomes even more  evident, further supporting their interpretation as   Bloch oscillations. 
Note  that the  two curves corresponding to  the chaotic regime (small tilt) lie  very close together  and reach  the highest equilibration values. As the tilt increases and the   system exits  the chaotic regime  ($D\geq1.6J$), 
the equilibration value of  the entanglement entropy   decreases, making this  observable  a  
sensitive  indicator  of the transition from  chaos to  integrability. However, as we will discuss below,   the depth of the correlation hole provides an even sharper   probe of this transition, as seen by   comparing   panels (b) and (d) of Figure~\ref{fig:sps_Ds}.

Panel (b) shows,  as a function of the tilt and for four different system sizes, the ratio between  the correlation-hole depth, $|\frac{1}{S_{Pmin}}-PR|$,  and the dimension of the Hilbert space,  $\mathcal{D}_\text{GOE}=\mathcal{D}/3$, while  panel (d)  presents the same but for  the equilibration value of the entanglement entropy  normalized  by the respective Page value.  
A first key observation of   both  panels  is that,  after the appropriate rescaling ---by the Hilbert‑space dimension for the correlation‑hole depth and by the Page value for the entanglement entropy—--the curves collapse onto nearly identical shapes across system sizes. This collapse indicates universal behavior. 
Likewise, both panels highlight that    the depth of the correlation hole is a significantly more  sensitive indicator  of the transition from chaotic to  regular dynamics: it drops sharply  from a finite value to nearly  zero around $D/J\approx 1.6$, in agreement  with the analysis based on  the mean level-spacing ratio in Fig.\ref{fig:rcaos}. In contrast, the equilibration value of the entanglement entropy  also decreases with increasing tilt, but with a much smoother trend. It is also worth noting that  in  the most chaotic case ($D=0.01J$),  which exhibits the deepest correlation hole in  panel (b), the  equilibrium value of the entanglement entropy    saturates at Page value, given  analytically   in Eq.~(\ref{eq:TheoreticalPV}). 


Panels (e) and (f) of Fig.~\ref{fig:sps_Ds}   show the corresponding  results for the imbalance. Panel (e) displays  its dynamics,  averaged over  Fock  states with initial   half-chain imbalance $\mathcal{I}_o=-1$. Panel (f) displays the equilibrium values as a function of  tilt   for four different system sizes. The  imbalance dynamics resembles that of the entanglement entropy: an initial rapid growth followed by oscillations, although   in this case  oscillations already  appear at small tilt. For the smallest tilt values, the imbalance equilibrates   close to zero,  $\mathcal{I}(t_{relax})=0$. As the tilt increases, the equilibrium value decreases significantly, reaching $\mathcal{I}(t_{relax})\approx -0.5$  for $D=0.8$. 
This value continues to decrease as the tilt grows and  the system crosses into the regular regime.  For large tilt, the  equilibrium imbalance   remains close to its initial value, approaching $\mathcal{I}(t_{relax})\approx -1$,     characteristic of Wannier-Stark localization.   
%
Thus, the decrease  in  the equilibrium   imbalance reflects the transition from a chaotic regime at small tilt to the regular  Wannier-Stark Localization regime at large tilt.


\begin{figure}
    \includegraphics[width=0.48\linewidth]{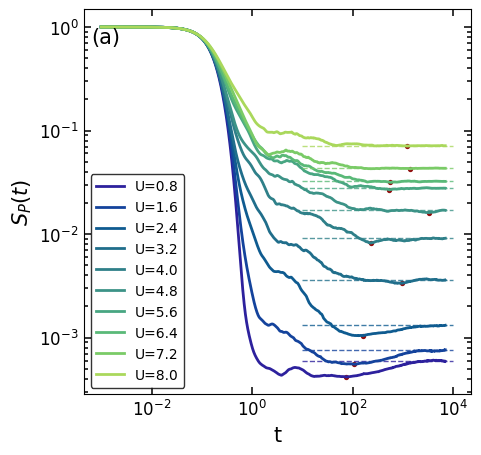}
    \includegraphics[width=0.48\linewidth]{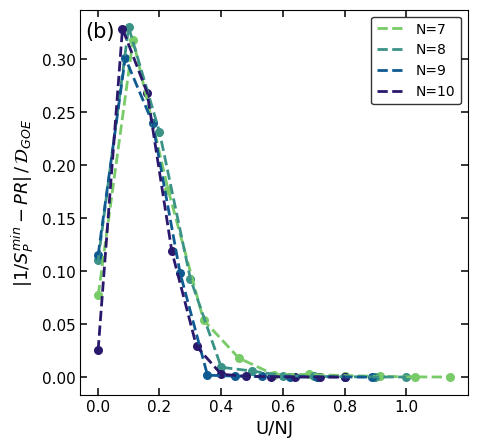}
    \includegraphics[width=0.48\linewidth]{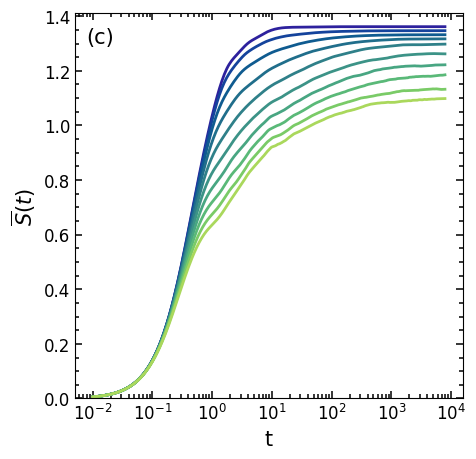}
    \includegraphics[width=0.48\linewidth]{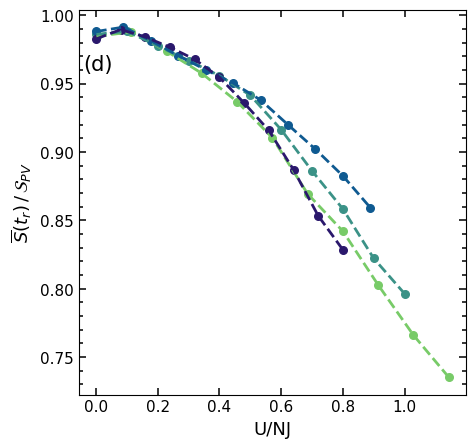}
    \includegraphics[width=0.48\linewidth]{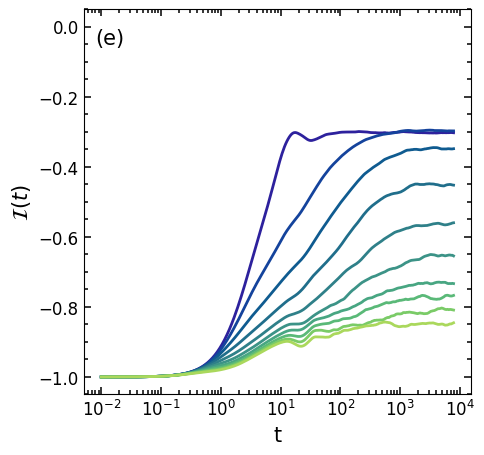}
    \includegraphics[width=0.48\linewidth]{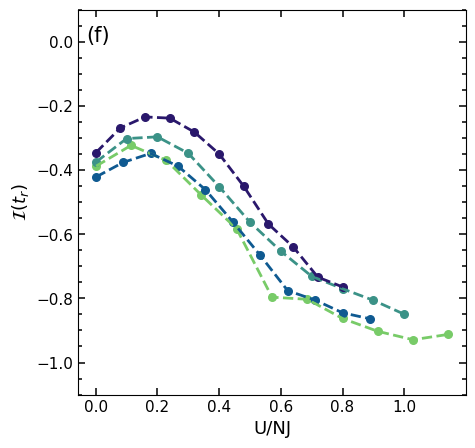}
    \caption{Same observables as Fig. \ref{fig:sps_Ds} but changing the interaction $U$,  with $D/J=0.5$}
    \label{fig:sps_Us}
\end{figure}

\subsubsection{Dynamics under interaction variation}

Figure~\ref{fig:sps_Us} presents  the results obtained when    varying  the interaction strength at fixed tilt $D=0.5J$.
In all cases, the survival probability in panel (a)  displays an initial quadratic decay which ---unlike the tilt-varying case --- is not followed by strong oscillations.  
From weak to intermediate interactions ($U\in [0.5,2.4]$), a clear correlation hole is visible  in the first three curves.
As $U/J$ increases, the correlation-hole  gradually disappears,   and the survival probability decays directly towards its relaxation value,  set  by the IPR. 

Panel (c) shows the  dynamics of the averaged  one-site entanglement entropy.  For chaotic cases,  the entropy grows rapidly and  reaches  a high  equilibrium value.  As $U$  increases and the system crosses to the regular  hard-core bosons regime, the equilibration value decreases and the equilibration time grows exponentially.  
The imbalance dynamics in panel (e) shows similar behavior:  chaotic cases reach larger equilibration values at significantly shorter times, whereas stronger interactions lead  to slower relaxation and lower  stationary imbalance.

The right panels of Fig.\ref{fig:sps_Us} show the depth of the correlation hole and the  equilibrium values of the  entanglement entropy and imbalance as  functions of the interaction strength  for different system sizes. As before, once   observables are properly  rescaled,  the curves  for all four system sizes collapse onto very similar trends. 
The interaction strength on the horizontal axes is scaled by the number of particles, consistent with the two-body character  of the interaction term.

Note that in the right  panels, unlike the left ones,  we also include results for  a very small interaction  ($U/J=0.01$), corresponding  to the  regular  Wannier-Stark localization regime. This is reflected in the fact that  the curves for the three observables  exhibit a maximum at  $U/(NJ)\sim 0.15$. 
Once again, the  depth of the  correlation-hole provides the sharpest and  more reliable indicator of the chaotic regime. Although  the equilibrium  entanglement entropy and  imbalance  also capture the transition, their variations are noticeably  smoother. 
For all  system sizes, the entanglement entropy reaches its maximum ---the  Page value--- at the same interaction strength where the correlation hole is deepest. However,  its decrease  is much more gradual than   the sharp drop observed in the  correlation-hole depth. In contrast,   the equilibrium imbalance exhibits  a steeper    decline  than the entanglement entropy, but  its maximum  occurs at slightly  larger  interaction  strength  than the one at which the  correlation-hole depth is maximal.


\section{Conclusions}\label{sec:conclusions}

In this work we investigated  the one-dimensional tilted Bose--Hubbard model with open boundary conditions and showed  that the competition between  the interaction term  and  the linear tilt generates a  rich dynamical landscape spanning  ergodic quantum chaos and integrable, nor-ergodic behavior. Two prominent limits structure this landscape:  the Wannier-Stark localization (WSL) regime  and the hard-core bosons (HCB) regime. On the static side, spectral diagnostics based on the mean ratio of consecutive level-spacings  delineate the crossover  between chaotic and regular regions across parameter space.  This spectral  picture   is corroborated by eigenstate-based  indicators,  including the participation ratio in the Fock basis and the one-site-averaged entanglement entropy, which display  clear qualitative changes  when moving between chaotic domains (characterized by relatively uniform eigenstate properties) and regular domains (where eigenstate indicators become strongly dispersed and structured). 

On the dynamical side, we found  that the survival probability $S_{P}(t)$ --- and, in particular,   the depth  of its correlation hole---  provides the sharpest and most direct dynamical witness of chaos throughout both crossovers, i.e., from chaos towards WSL and from chaos towards HCB. In the chaotic regime, $S_{P}(t)$ exhibits  the characteristic dip--ramp--plateau behavior associated with long-range spectral correlation,  and   the  correlation-hole depth follows the expected scaling with Hilbert-space size, consistent with  GOE expectations. 
 As the system becomes regular, the correlation hole disappears, yielding a clean dynamical signature of the loss of spectral rigidity.

These conclusions are broadly consistent with the behavior of two experimentally accessible observables: the on‑site entanglement entropy and the half‑chain imbalance. In the tilt‑dominated (WSL) regime ($D/J~\gg~U/J$), the dynamics is marked by pronounced oscillations consistent with Bloch oscillations, the absence of a correlation hole in $S_{P}(t)$, reduced long‑time entanglement, and an effective freezing of the imbalance near its initial value. In the interaction‑dominated (HCB) limit ( $U/J~\gg~D/J$), Bloch oscillations are absent, but the remaining signatures of regular dynamics persist: relaxation slows down, entanglement equilibrates to lower values, and the imbalance retains strong memory of the initial condition. By contrast, when tilt and interactions are comparable, the system enters an extended chaotic regime in which the correlation hole becomes visible, the entanglement entropy rapidly approaches the analytically available  Page value, and the imbalance relaxes toward values close to zero, consistent with efficient redistribution of particles across the chain.

Importantly, the equilibrium (long‑time) values of all three dynamical probes track the regular‑to‑chaotic crossover and—once appropriately rescaled—exhibit universal trends that are largely independent of system size. The entanglement entropy, in particular, becomes naturally comparable across sizes when normalized by the corresponding Page value, revealing a robust scaling collapse. Among all observables considered, the survival probability remains the most reliable and sensitive indicator of chaos, because it directly encodes long‑range spectral correlations through the correlation hole. Interestingly, we also find that the imbalance 
provide a sharper contrast than entanglement entropy across parts of the crossover, despite entanglement being the more standard diagnostic of thermalization.

Overall, our results establish the correlation hole in the survival probability as a stringent dynamical marker of chaos in the tilted Bose–Hubbard model, while entanglement and imbalance provide complementary, experimentally accessible signatures. The observed scaling collapses suggest a universal description of the chaos–regular crossover that persists across system sizes and particle numbers.

\section*{Acknowledgments}
\label{sec:Acknowledgments}

We acknowledge the support of the Computation Center - ICN, in particular to Enrique Palacios, Luciano D\'iaz, and Eduardo Murrieta. C.D-M. and J.G.H acknowledge partial financial support from the DGAPA-UNAM projects IN109523 and IN101526. 

\appendix
\section{
Set of initial Fock states and scaling of the chaos transition}
\label{sec:apendix1}

In this appendix, we provide additional  details on 
the set of initial Fock  states used to study the dynamics of entanglement entropy and imbalance, illustrated  for the case $N=M=8$.

\begin{figure}[h]
    \centering
    \includegraphics[width=1\linewidth]{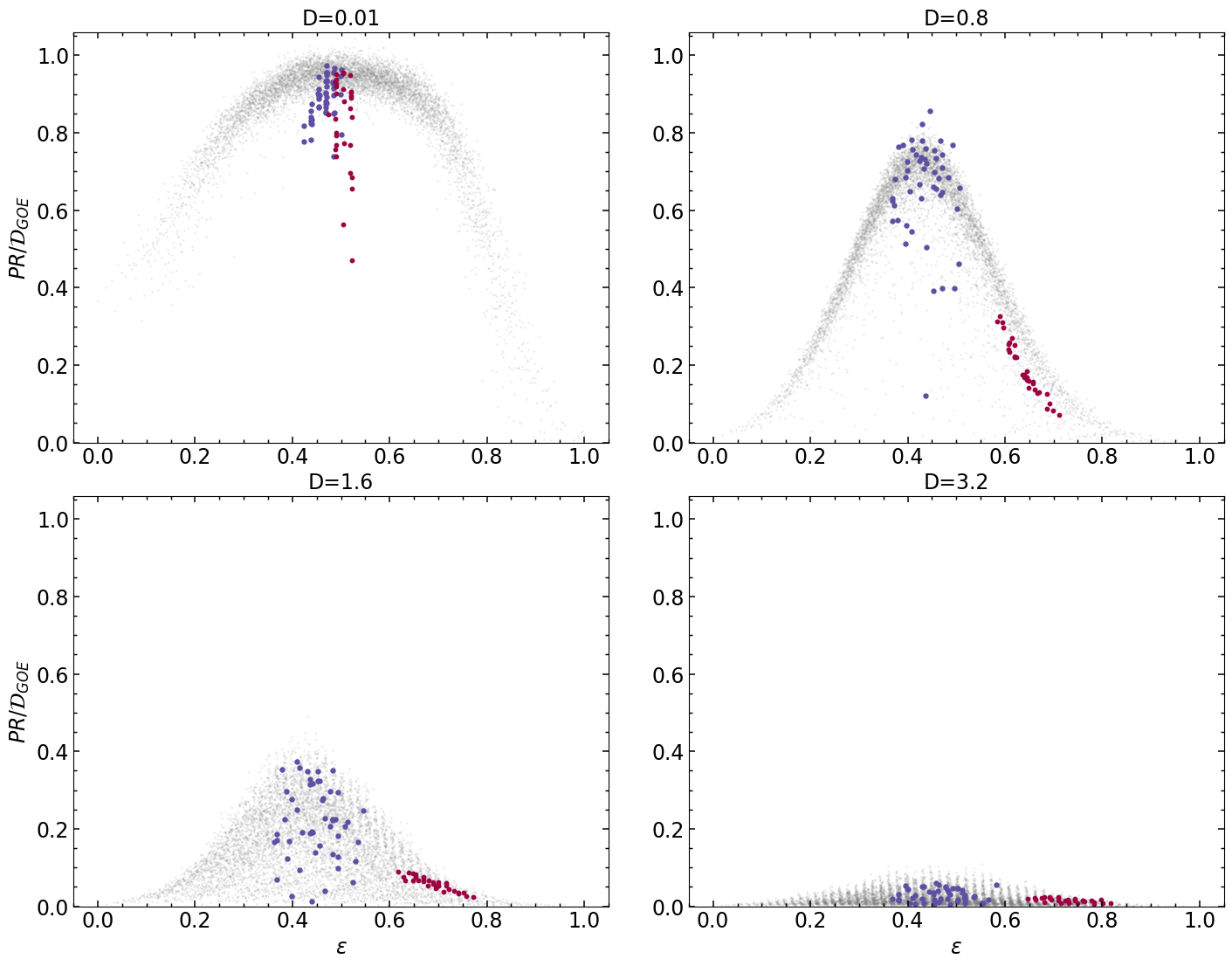}
    \caption{Distribution of the Participation ratio (PR) of eigenstates in the Fock basis (light gray background) for different tilt values  at fixed interaction $U/J=0.5$, with $N=M=8$, as a function of their normalized eigenenergies $\epsilon$. Blue dots indicate the mean energy and participation ratio in the energy eigenbasis of the    50 random initial Fock states used to study the dynamics of  entanglement entropy. 
    Red dots represent  correspond to the set of 31 initial states with fixed initial imbalance $\mathcal{I}(0)=-1$. All states are restricted to a maximum site occupation of $n_{i}\leq 3$ bosons. }
    \label{fig:Prs_Fock1}
\end{figure}

\begin{figure}[h]
    \centering
    \includegraphics[width=1\linewidth]{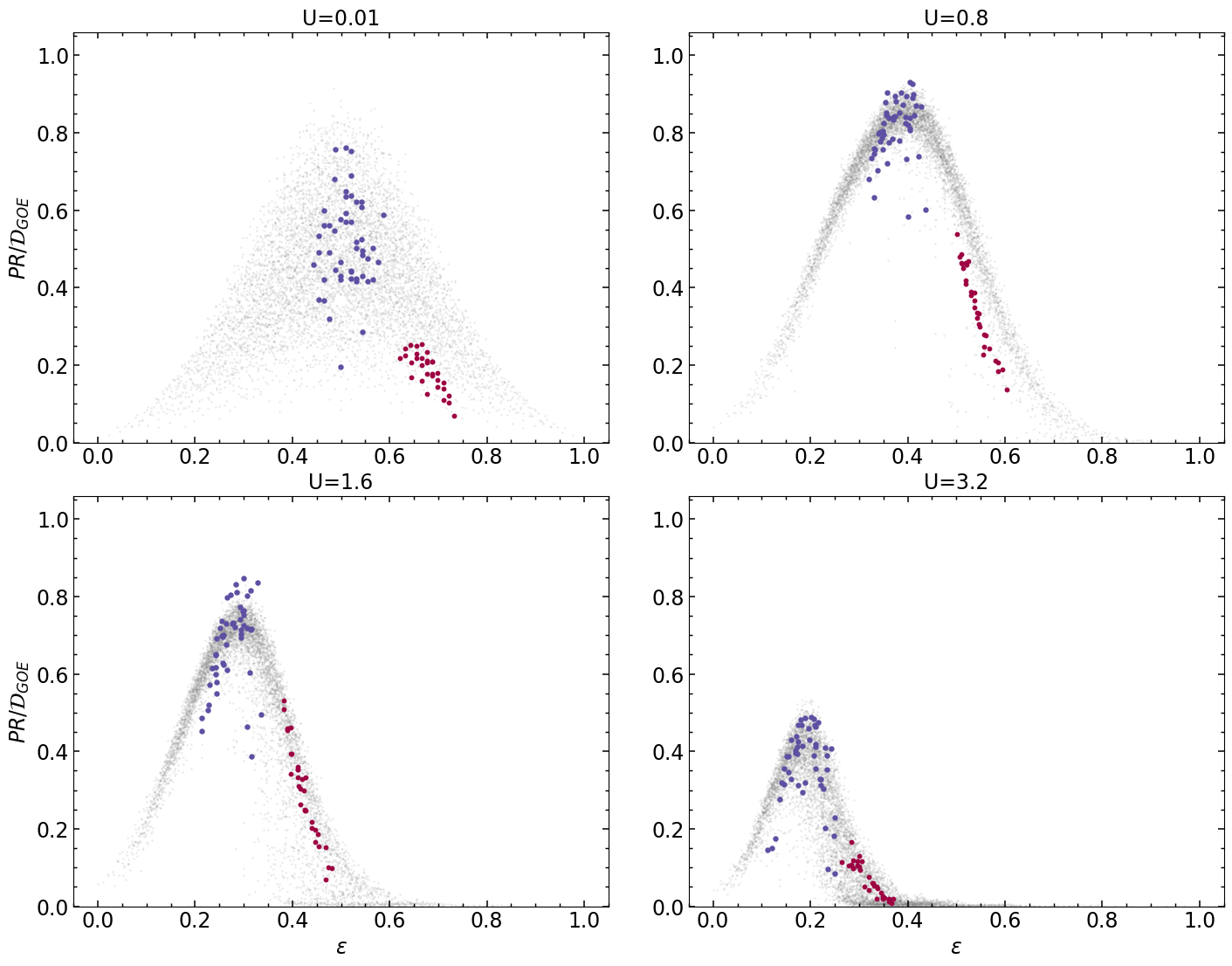}
    \caption{Same analysis as in Fig. \ref{fig:Prs_Fock1}, but varying the interaction strength at a fixed tilt $D/J=0.5$. }
    \label{fig:Prs_Fock2}
\end{figure}

Figures \ref{fig:Prs_Fock1} and \ref{fig:Prs_Fock2} show, for varying tilt and interaction strength,   the mean energy and participation ratio in the Hamiltonian eigenbasis  of the selected  initial Fock  states. The states  used to study  entanglement entropy are marked with blue dots across  all panels, while   those used for  imbalance are indicated with red dots. For  reference,   the participation ratio of the respective  Hamiltonian eigenstates in the Fock basis is plotted against their rescaled energies using gray dots. 

The initial Fock states for  entanglement entropy were chosen as 50 random states located   near the center of the spectrum  for $U=0.5 J$ and $D=0.8 J$. This explains why the blue dots in the top-right panel of  Fig. \ref{fig:Prs_Fock1} cluster around  the middle-energy region. In the other panels, the blue dots spread  across the energy range,  though they remain close to the  region where  Hamiltonian eigenstates exhibit larger participation ratios.      

By contrast,  states with a fixed initial imbalance of $\mathcal{I}(0)=-1$ tend to shift towards the high-energy edge of the spectrum. As a result, these states undergo localization in the Hamiltonian eigenbasis more rapidly than those selected from the spectral center. This spectral migration accounts for the faster  decay of imbalance relaxation values compared  to  those of the entanglement entropy as tilt or interaction strength increases.

\begin{figure}[h]
    \centering
    \includegraphics[width=0.49\linewidth]{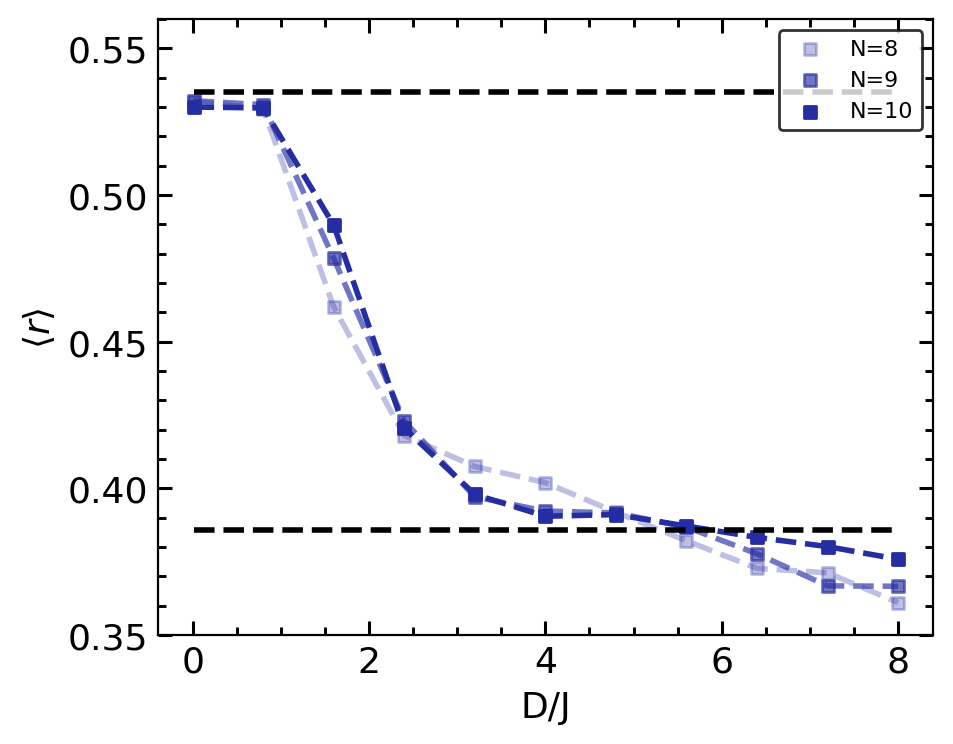}
    \includegraphics[width=0.49\linewidth]{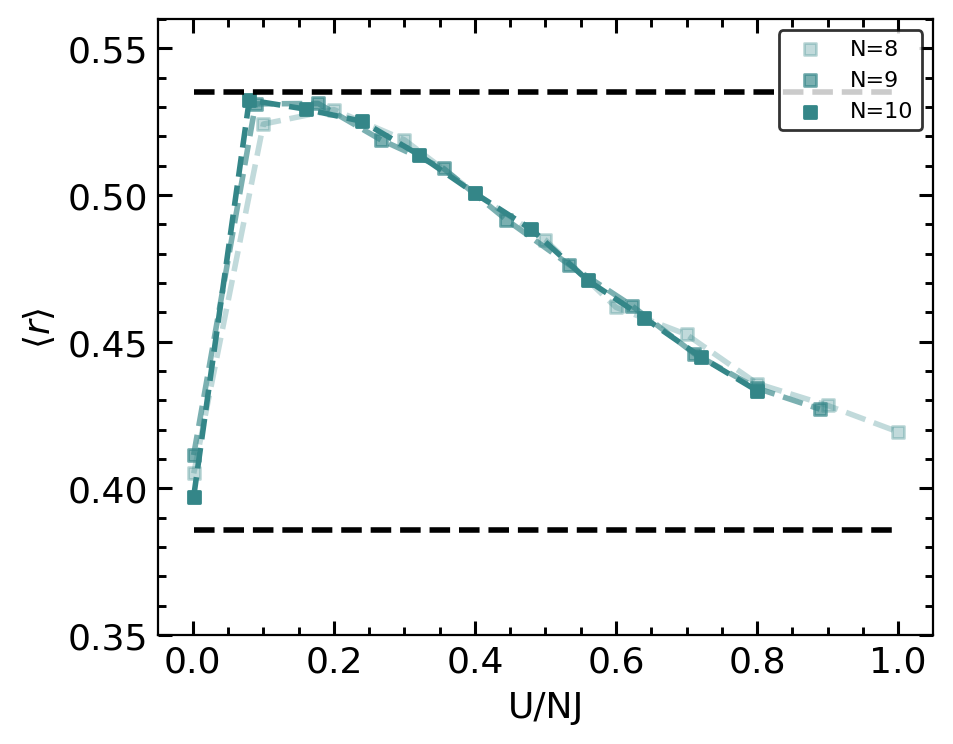}
    \caption{Mean level spacing ratio $\langle r \rangle$ for different system sizes $N \in [8, 9, 10]$. The green curves (right panel) represent the change in interaction $U$ (at fixed tilt), while the blue curves (left panel) represent the change in tilt $D$ (at fixed interaction). The upper and lower black dashed lines indicate the theoretical values for the GOE ($\langle r \rangle_{\text{GOE}} \approx 0.536$) and Poisson ($\langle r \rangle_{\text{P}} \approx 0.386$) limits, respectively.}
    \label{fig:rcaos_U_D}
\end{figure}

Finally, in Fig. \ref{fig:rcaos_U_D}, we present a finite-size scaling analysis of the mean level spacing ratio to test the robustness of the  transition between  regular and chaotic regimes against  changes in system  size.  This analysis supports the static phase diagram shown  in Fig. \ref{fig:rcaos}, revealing   no significant dependence on system size  within the studied range $N \in [8, 9, 10]$. It is worth noting that, consistent with  other similar  scaling analyses   in the main text,   the  dependence of $\langle r \rangle$ on the interaction strength (right panel) was    normalized  by the number of particles $N$, in agreement with the two-body nature of this interaction term.     


\bibliographystyle{unsrt}
\bibliography{biblio2}
\end{document}